\documentstyle[12pt]{report}

\catcode`\^^?=9 

\def\half{\mbox{\small{$\frac{1}{2}$}}}
\def\third{\mbox{\small{$\frac{1}{3}$}}}
\def\quarter{\mbox{\small{$\frac{1}{4}$}}}
\def\mfrac#1#2{\mbox{\small{$\frac{#1}{#2}$}}}

\def\be{\begin{equation}}
\def\ee{\end{equation}}
\def\ba{\begin{eqnarray}}
\def\ea{\end{eqnarray}}
\def\mref#1{eq.(\ref{eq:#1})}
\def\nref#1{(\ref{eq:#1})}
\def\mlab#1{\label{eq:#1}}

\def\mref#1{Eq.(\ref{eq:#1})}

\begin{document}

\begin{center}
{\LARGE Chiral Symmetry Breaking with the Curtis-Pennington Vertex}
\vskip1cm

{\bf D. Atkinson, V.P. Gusynin\footnote{Permanent address: Institute
for Theoretical Physics, Ukrainian Academy of Sciences,
252143 Kiev, Ukraine.}
and P. Maris

Institute for Theoretical Physics

Nijenborgh 4

9747 AG Groningen, The Netherlands

{\tiny \bf E-mail: ATKINSON@TH.RUG.NL}}

\vskip4cm

{\parbox{12cm}{{\small \bf Abstract:} \noindent \small \sf
We study chiral symmetry breaking in quenched
QED$_4$, using a vertex Ansatz recently proposed by Curtis and
Pennington.  Bifurcation analysis is employed to establish the existence of
a critical coupling and to estimate its value.
The main results are in qualitative agreement with the
ladder approximation, the numerical changes being minor.}}
\end{center}
\vskip1cm
\pagestyle{empty}

\chapter{Introduction}

The study of chiral symmetry breaking in quantum field theory requires
the use of non-perturbative techniques. A commonly used method is based on
solving the Dyson-Schwinger equations in the so-called rainbow or ladder
approximation, which uses a bare
vertex\cite{Maskawa},\cite{Valery},\cite{Valery1},\cite{Bardeen}.
However, such a truncation of
the infinite set of integral equations in QED does not in general respect
two fundamental requirements: the Ward-Takahashi identity and
multiplicative renormalizability. This shortcoming calls in question all
the results of the ladder approximation.

Recently Curtis and Pennington\cite{CP}  presented a very interesting
Ansatz for the full
three-point vertex that satisfies the Ward-Takahashi identity and which is
multiplicatively renormalizable to all orders in leading and
next-to-leading logarithms in perturbation theory.
They showed how powerful  the
constraints on the form of the vertex are. The natural question is to
what extent the Curtis-Pennington modification to the ladder
approximation alters the physical results, in particular the value of
the critical coupling for onset of chiral symmetry breaking.
Applying bifurcation theory to the Curtis-Pennington equations, we
find that there is indeed a non-zero critical coupling; moreover the
numerical changes as compared with the ladder approximation will be shown to be
fairly small.

These relatively minor numerical modifications encourage the hope that the
essential results of the ladder approximation survive vertex
modifications.  However, caution is necessary in espousing
the point of view that the full theory has the same properties,
since it is known\cite{Kondo} that vacuum polarization
effects {\em do modify} the nature of the chiral-symmetry phase
transition.  Indeed, mean field behaviour appears to supervene.
Nevertheless, the mutual influence of two essentially distinct but
related effects, that of charge screening and of `fall into the centre',
have not yet been consistently taken into account. The basic question:
`Does the ladder approximation already contain the essence of the
physics of chiral symmetry breaking?' is therefore still open.

In this paper we shall neglect all vacuum polarization effects and we
restrict attention to the Landau gauge. Moreover we neglect another
known deficiency of the ladder approximation, namely the occurrence of
spurious first-sheet complex singularities in the fermion
propagator\cite{Blatt}. We write the equations from the beginning in
Euclidean space and postpone the question of their relevance to
Lorentz space.
\pagestyle{plain}

\chapter{Curtis-Pennington Equations}
In Euclidean space our notation for the unrenormalized fermion
propagator is\footnote{Curtis and Pennington's $\Sigma$ is our $m$ and
their $F$ is our $A^{-1}$.}
\be
S_F(p)=\frac{A^{-1}(p^2)}{\gamma p+m(p^2)}\, .
\ee
The Landau gauge bare photon propagator has the form
\be
D^{\mu\nu}_F (k) = \frac{1}{k^2}
\left( \delta^{\mu\nu}-\frac{k^{\mu}k^{\nu}}{k^2}\right)\, ,
\ee
and the Curtis-Pennington equations are
\be
A(x)=1-\frac{\alpha_0}{4\pi}\int_0^{\Lambda^2}\frac{dy}{y+m^2(y)}I(y,x)
\mlab{CP1}
\ee
and
\be
A(x)m(x)=\frac{\alpha_0}{4\pi}\int_0^{\Lambda^2}
\frac{dy}{y+m^2(y)}J(y,x)\, .  \mlab{CP2}
\ee
The functions $I$ and $J$, which depend explicitly on $m$ and $A$, were
given in Ref.\cite{CP}. If we set $m(x)=0$, we find that $I$
and $J$ both vanish, in fact $I$ is of order $m^2$ while $J$ is of order
$m$. Hence \mref{CP1} and \mref{CP2} possess the trivial solution
$A(x)=1$ and $m(x)=0$. This is the only solution for small values of the
coupling, $\alpha_0$.

In order to examine the possibility that a nontrivial solution, $m(x)$,
branches away from the trivial one at a critical coupling, $\alpha_c$,
we examine the so-called bifurcation equation\cite{Pimbley}. This
involves differentiating
\mref{CP1} and \mref{CP2} functionally with respect to $m$ and then
setting $m=0$. Since $I$ is second-order in $m^2$, we find to first
order in $\delta m$ that $I(y,x)= 0 $ and, from the
Curtis-Pennington expressions,
\ba
J(y,x) &=&3\delta m(y)
\left[ \frac{y}{x}\theta (x-y) +\theta (y-x) \right]
\nonumber \\
&&-\frac{3x}{2} \frac{\delta m(y)-\delta m(x)}{y-x}
\left[ \frac{y^2}{x^2}\theta (x-y) +\theta (y-x) \right]
\ea

According to the standard rules of bifurcation theory,
we should substitute $y$ for the denominator
${y+m^2(y)}$ in  \mref{CP2}. However, this introduces a spurious
infra-red singularity, which can be eliminated by replacing the
denominator by $y+m^2$, where the constant $m$ can be fixed {\em a
posteriori} by the normalization requirement
\be
\delta m(0)=m\, . \mlab{IR}
\ee
We shall call $m$ the dynamical mass.
Rescaling $x/m^2\longrightarrow x$,
$y/m^2\longrightarrow y$, and $\delta m(x)/m \longrightarrow f(x)$,
we obtain the basic bifurcation equation
\ba
f(x)&=&\frac{\lambda}{x} \int_0^x \frac{y dy}{y+1}\left[
f(y)-\frac{y}{2}\frac{f(y)-f(x)}{y-x}
\right]
\nonumber \\
&&+\lambda \int_x^\frac{\Lambda^2}{m^2} \frac{dy}{y+1}\left[
f(y)-\frac{x}{2}\frac{f(y)-f(x)}{y-x}\right]\, ,
\mlab{furc}
\ea
where
\[
\lambda =\frac{3\alpha_0}{4\pi}\, .
\]
In the Landau gauge,
the terms involving $\frac{f(y)-f(x)}{y-x}$ are all that is left of the
complicated Curtis-Pennington Ansatz in the linearized form appropriate
to a bifurcation analysis.

\chapter{Critical Coupling}

In this section we shall consider three approximate methods for
handling the bifurcation equation \nref{furc}:
\begin{enumerate}
\item $\frac{f(y)-f(x)}{y-x} \Longrightarrow 0$
\item $\frac{f(y)-f(x)}{y-x} \Longrightarrow f'(y)$
\item $\frac{f(y)-f(x)}{y-x} \Longrightarrow f' \left(
{\mbox{\small \sf max}}{ (x,y) }\right)$\, .
\end{enumerate}

\section{Second-order Equation}

Here we simply ignore the extra Curtis-Pennington term. As is
well-known, the integral equation
\be
f(x)=\frac{\lambda}{x} \int_0^x \frac{y dy}{y+1}f(y)
+\lambda \int_x^\frac{\Lambda^2}{m^2} \frac{dy}{y+1}
f(y)\, ,
\mlab{furc1}
\ee
then reduces to the standard
hypergeometric differential equation
\be
x(x+1)f''+2(x+1)f'+\lambda f=0\, ,
\ee
with infra-red and ultra-violet boundary
conditions.
The required solution is
\be
\mlab{F21}
f(x)= {_2F_1}(\alpha_1,\alpha_2;\gamma ;-x) \, ,
\ee
where
\be
\alpha_1, \, \alpha_2 =\half\pm\sqrt{\lambda_c-\lambda}\, ;\hspace{5mm}
\gamma=2\, ;\hspace{5mm}
\lambda_c=\quarter \, .
\ee
This solution of the differential equation is the only one that is
finite at $x=0$: the normalization is fixed by the infra-red condition
$f(0)=1$, which corresponds to \mref{IR}. From the integral equation we
still have to impose the ultra-violet condition,
$B(\frac{\Lambda^2}{m^2})=0$, where
\be
B(x)= \frac{d}{dx}\left[ xf(x)\right] =
{_2F_1}(\alpha_1 ,\alpha_2;1;-x) \, .   \mlab{(xm)'}
\ee
For $\lambda < \lambda_c$, $\alpha_1$ and $\alpha_2$
are real, and under these conditions
the hypergeometric function \mref{(xm)'} has no finite zeros: the
ultra-violet condition cannot be satisfied for any $m \neq 0$, so
there is no nontrivial solution of the integral equation. The imposition
of the ultra-violet cut-off, $\Lambda$, is essential for this
result\cite{Maskawa},\cite{Valery},\cite{Valery2},\cite{Fred}.
{}From the physical point of view, it is easy to satisfy the
axial-vector current conservation with a finite cut-off by just setting
the bare mass equal to zero. Without an ultra-violet
cut-off, one needs a more subtle treatment, distinguishing between
an irregular solution, which exists for any coupling strength, but which
leads to non-conservation of the axial-vector current, and the regular
solution, which exists only if the coupling is large enough. It is not
easy in a computer calculation to see the difference between these two
types of solution unless one explicitly works with a
cut-off\footnote{In the second paper of
Ref.\cite{CP} Curtis and Pennington had
claimed that  a dynamical mass is generated for all values of the
coupling. In fact they failed to make the above distinction between a
regular and an irregular solution.}. A
cut-off is also necessary in order to make sure that the bifurcation equation
is of classical Fredholm type, so that the standard theory can be
applied\cite{Pimbley}.

When $\lambda > \lambda_c$, $\alpha_1$  and $\alpha_2$
are complex conjugate numbers and
the hypergeometric function \mref{(xm)'}, which remains real, has
an infinite number of oscillations; the smallest zero, say $x_0$, can be
chosen to define $m$ through the relation
\[
x_0=\frac{\Lambda^2}{m^2}\, .
\]
For large $\Lambda$ , the dynamical mass is given by
\be
\mlab{m2}
m=\Lambda
\exp \left[ -\frac{\pi + \delta (\lambda )}{2\sqrt{\lambda -\lambda_c}}\right]
\ee
with
\be
\mlab{m22}
\delta (\lambda )=  \arg \frac
{\Gamma^2 (\frac{1}{2}+i\sqrt{\lambda -\lambda_c})}
{\Gamma (1+2i\sqrt{\lambda -\lambda_c})}
\, .
\ee

Near the critical point, $\lambda=\lambda_c=0.25$, \mref{m2} takes the
form
\be
m=\Lambda \exp \left[
-\frac{\pi /2}{\sqrt{\lambda -\lambda_c}}+2\log 2 \right] \, ,
\mlab{mag2}
\ee
whereas the mass function \nref{F21} behaves at large momenta as
\be
\mlab{logas}
f(x)\sim \frac{1}{\sqrt{x}}\log x \, .
\ee
The formulas \nref{m2} and \nref{m22}
have played a great role in attracting attention
to the non-perturbative study of QED. Firstly the critical point
$\alpha_c=\frac{\pi}{3}$ has been
interpreted \cite{Valery,Valery1}
as a nontrivial ultra-violet fixed point defining the
continuum limit of QED. In fact, if we start from the supercritical phase
and let $\Lambda \rightarrow \infty$,
then in order to keep the dynamical mass finite we have to perform the
 charge renormalization as follows
\[
\alpha (\Lambda )=\alpha_c \left( 1+\frac{\pi^2}{\log^2
\frac{4\Lambda}{m}}\right)
\longrightarrow \alpha_c =\frac{\pi}{3}\, ,
\]
as $\Lambda \rightarrow \infty$.
Furthermore, the $\beta$-function of the
theory,
\[
\beta (\alpha )\equiv \Lambda \frac{\partial \alpha}{\partial \Lambda}
=-\frac{2}{3} \left( \frac{\alpha}{\alpha_c} -1\right)^{\frac{3}{2}},\,
 \hspace{6mm}\alpha >\alpha_c,
\]
has an ultra-violet stable zero at $\alpha =\alpha_c$.
The existence of such a critical
point was confirmed recently by computer simulations of
noncompact QED on the lattice\cite{Kogut}.
Secondly, as one can see from \mref{logas}, the fermion
dynamical mass falls off with increasing momentum as $\frac{1}{p}$,
and this is in
sharp contrast to asymptotically free theories like QCD where
$m(p^2)\sim \frac{1}{p^2}$ \cite{Lane}.
It was suggested\cite{Yamawaki} that this more slowly decreasing
behaviour of the mass-function can
solve the notorious flavour-changing neutral current
problem in extended technicolour theories.

By proceeding beyond the
ladder approximation, we would like to find out whether the results
\nref{m2} and \nref{m22}
are still valid or not.
In Ref.\cite{Holdom},\cite{Mahanta} evidence was adduced
in favour of the validity of
an equation of the type \nref{m2}
at any finite order in the quenched approximation, and we
shall make use of some of the arguments in Sect. 4. Our main concern here will
be to take into account the vertex corrections in a manner compatible
with the requirements of both the Ward-Takahashi identity and
of multiplicative renormalizability.

\section{Third-order Equation}

With the second approximation, the integral equation reads
\ba
f(x)&=&\frac{\lambda}{x} \int_0^x \frac{y dy}{y+1}\left[
f(y)-\frac{y}{2}f'(y)
\right]
\nonumber \\
&&+\lambda \int_x^\frac{\Lambda^2}{m^2} \frac{dy}{y+1}\left[
f(y)-\frac{x}{2}f'(y)\right]\, ,
\mlab{furc2}
\ea
We shall show that this integral equation is equivalent to a third-order
differential equation of generalized hypergeometric type, together with
an infra-red and {\em two} ultra-violet boundary conditions.

A first differentiation of \mref{furc2} gives
\ba
f'(x)&=&-\frac{\lambda}{x^2} \int_0^x \frac{y dy}{y+1}\left[
f(y)-\frac{y}{2}f'(y)
\right]
\nonumber \\
&&-\frac{\lambda}{2} \int_x^\frac{\Lambda^2}{m^2}
\frac{dy}{y+1}f'(y) \, ,
\mlab{furc21}
\ea
which implies the infra-red condition
\be
{\left. x^2f'(x)\right|}_{x=0}=0\, .
\ee
On combining \mref{furc2} and \mref{furc21} we find
\be
{\left[ xf(x)\right]}' =
\lambda\int_x^\frac{\Lambda^2}{m^2}
\frac{dy}{y+1}\left[ f(y)-xf'(y)\right] \, .\mlab{furc22}
\ee
A second  differentiation gives
\be
{\left[ xf(x)\right]}'' = \frac{\lambda}{x+1}\left[ xf'(x)-f(x)\right] -
\lambda\int_x^\frac{\Lambda^2}{m^2}
\frac{dy}{y+1}f'(y) \, ,\mlab{furc23}
\ee
and a third differentiation finally yields
\be
f''' +\left[ \frac{3}{x}-\frac{\lambda}{x+1}\right] f''
-\frac{\lambda}{x(x+1)^2}f'-\frac{\lambda}{x(x+1)^2}f=0\, .
\mlab{furc24}
\ee
The first ultra-violet condition (from \mref{furc22}),
\be
\left. {\left\{ xf(x)\right\} }'\right|_{x=\frac{\Lambda^2}{m^2}} = 0\, ,
\ee
is as in the second-order case; but \mref{furc23} gives rise to a new
ultra-violet condition,
\be
\left. \left\{ [xf(x)]''+\lambda\frac{f(x)-xf'(x)}{x+1}\right\}\right|_{
x=\frac{\Lambda^2}{m^2}} = 0\, .
\ee

The above differential equation belongs to the class of
equations of Fuchsian type with three regular singularities at the points
$x=-1,0,\infty$.
Moreover, changing the variable to $z=x+1$ and writing \mref{furc24} as
\be
z^2(1-z)f''' +\left[ -\lambda -(3-\lambda )z\right] zf''
+\lambda f'+\lambda f=0\, , \mlab{hyper1}
\ee
we recognize it as the equation for the hypergeometric function,
${_3F_2}$.

\begin{picture}(0,0)(1,22)
\put(74,15){$\scriptstyle
\alpha_i+\gamma_1 ,\, \alpha_i+\gamma_2 ,\, \alpha_i+\gamma_3$}
\put(74,10){$\scriptstyle \alpha_i-\alpha_1+1 , \alpha_i-\alpha_2 +1 ,
\alpha_i-\alpha_3+1$}
\put(116,10){\line(0,1){7}}
\put(118,14){$1$}
\put(118,12){--}
\put(118,10){$z$}
\end{picture}
We choose as a linearly independent set of solutions
\be
f_i(z)=\left\{ \prod_{\nu =1}^3 \frac{\Gamma (\alpha_i+\gamma_\nu )}
{\Gamma (\alpha_i-\alpha_\nu +1)}\right\} z^{-\alpha_i}
{_3F_2}\left(
\begin{array}{l}
\phantom{\alpha_i+\gamma_1 ,\, \alpha_i+\gamma_2 ,\hspace{7mm}} \\
\phantom{\alpha_i-\alpha_1+1 , \alpha_i-\alpha_2 ,\hspace{7mm} }
\end{array}
\right)  \mlab{fi}
\ee
with the convention that the term $\alpha_i-\alpha_j+1$
in ${_3F_2}$
is omitted when
$i=j$. Here the $\alpha_i$ are the roots of the characteristic equation
\be
\mlab{char}
\alpha ^3+\lambda \alpha ^2+(\lambda -1)\alpha +\lambda =0\, ,
\ee
and $\gamma_1=0\,$, $\gamma_2=1+\lambda\,$, $\gamma_3=2\,$.
The $\alpha$'s (which should not be confused with the coupling,
$\alpha_0$!)
evidently satisfy
\[
\alpha_1+\alpha_2+\alpha_3 =-\lambda\, ,\hspace{1cm}
\alpha_1\alpha_2+\alpha_2\alpha_3+\alpha_3\alpha_1 =\lambda -1 \, ,\hspace{1cm}
\alpha_1\alpha_2\alpha_3 = -\lambda \, .
\]
The $\Gamma$ factors are included for later convenience.

The general solution of  \mref{hyper1} is given by an arbitrary linear
superposition of the $f_i$ defined in   \mref{fi}. Each of these
functions behaves like ${(1-z)}^{-1}$ as $z \rightarrow 1$. However,
N\o rlund\cite{Norlund} showed that the difference between any two of the
solutions \mref{fi} is regular at $z=1$. We require this regularity
because $z=1$ corresponds to $x=0$, and the infra-red condition imposes
finiteness of $f(1)$. We choose as two regular
solutions the following:
\ba
f_{1,3}(z)&=&f_3(z)-f_1(z)\nonumber\\
f_{2,3}(z)&=&f_3(z)-f_2(z)\, .\nonumber
\ea
These solutions, together with any one of   \mref{fi}, also form a
fundamental system. Thus the most general solution can be written in the
form
\[
f(z)=c_1f_{1,3}(z)+c_2f_{2,3}(z)+c_3f_3(z)\, .
\]
The infra-red condition immediately gives $c_3=0$, and the ultra-violet
conditions lead to a system of linear homogeneous equations for the
constants $c_1$ and $c_2$.
In order for there to be a nontrivial solution of these homogeneous
equations, the corresponding determinant must vanish, and one obtains in
this way  the ratio
$c_1/c_2$. The undetermined common constant is fixed by the
normalization $f(1)=1$. This vanishing of the determinant yields an
equation for the dynamical mass; considering the large $\Lambda$
limit, we find, after some algebra,
\ba
(\alpha_2-\alpha_3)(\alpha_2+\alpha_3-1)\alpha_1
&&\hspace{-6mm}\prod_{i=1}^3
\frac{\Gamma (\alpha_2+\gamma_i)\Gamma (\alpha_3+\gamma_i)}
{\Gamma (\alpha_2-\alpha_i+1)\Gamma (\alpha_3-\alpha_i+1)}
{\left( \frac{\Lambda^2}{m^2}\right)}^{-\alpha_2-\alpha_3}
\nonumber\\&&
\nonumber\\
&&\hspace{-5mm}+\mbox{\sf two cyclic permutations of $\{ 1,2,3\}$}=0
\, .
\mlab{algebra}
\ea

In the case that the $\alpha$'s are real, and
with the ordering $\alpha_3\le \alpha_2\le \alpha_1$, the term written
explicitly in \mref{algebra} is the leading one, and in this case there
is no nontrivial solution of
\mref{algebra}, and hence none of \mref{furc}. However, on increasing
$\lambda$ (or equivalently the coupling $\alpha_0$), two of the roots
($\alpha_1$ and $\alpha_2$ with our ordering)
become equal, and then complex conjugate. For a cubic equation, it is
easy to ascertain when this happens by looking at Cardano's formulas:
\[
\alpha_3=A+B-\frac{\lambda}{3}\, ,\hspace{1cm}
\alpha_1,\alpha_2=-\frac{A+B}{2}\pm \frac{\sqrt{3}}{2}i(A-B)-
\frac{\lambda}{3}\, ,
\]
where
\[
A={\left( -\frac{q}{2}+\sqrt{Q}\right)}^\frac{1}{3}\, ,
\hspace{1cm}
B={\left( -\frac{q}{2}-\sqrt{Q}\right)}^\frac{1}{3}\, ,
\hspace{1cm}
Q={\left(\frac{p}{3}\right)}^3+{\left(\frac{q}{2}\right)}^2 \, ,
\]
\[
p=-\third \lambda^2+\lambda -1\, , \hspace{1cm}
q=\mfrac{2}{27}\lambda^3-
\mfrac{1}{3}\lambda^2+
\mfrac{4}{3}\lambda\, .
\]

When $Q<0$, \mref{char} has three different real roots; but if $Q>0$
there are one real root and two complex conjugate ones. In the latter
case two terms in \mref{algebra} are of the same order, and then the
equation for the dynamical mass can be written in the form
\ba
&&\alpha_2(\alpha_3 -\alpha_1)(\alpha_3 +\alpha_1-1)
\prod_{i=1}^3\frac{\Gamma (\alpha_1+\gamma_i)}
{\Gamma (\alpha_1-\alpha_i+1)}
{\left( \frac{\Lambda^2}{m^2}\right)}^{-\alpha_1}
\nonumber\\ \nonumber
&&-
\alpha_1(\alpha_3 -\alpha_2)(\alpha_3 +\alpha_2-1)
\prod_{i=1}^3\frac{\Gamma (\alpha_2+\gamma_i)}
{\Gamma (\alpha_2-\alpha_i+1)}
{\left( \frac{\Lambda^2}{m^2}\right)}^{-\alpha_2}
=0 \,   .
\ea
Writing $\alpha_1=u+iv$ and $\alpha_2=u-iv$, we obtain
\[
\sin \left[ v \log \frac{\Lambda^2}{m^2}-\delta (\lambda )\right] =0 \, ,
\]
where the phase is given by
\[
\delta (\lambda )=\arg \left[
\alpha_2 (\alpha_3 -\alpha_1)(\alpha_3 +\alpha_1-1)
\prod_{i=1}^3\frac{\Gamma (\alpha_1+\gamma_i)}
{\Gamma (\alpha_1-\alpha_i+1)}
              \right]\, ,
\]
and the mass is
\[
m=\Lambda \exp \left[ -\frac{\pi n+\delta (\lambda )}{2v}\right]\, ,
\hspace{1cm}n=1,2, \ldots \, .
\]
As usual, the value $n=1$ corresponds to the ground state of the system. The
critical coupling constant $\lambda_c(=\frac{3\alpha_c}{4\pi})$,
separating the chirally symmetric phase from that of spontaneous chiral
symmetry breaking, is determined by the characteristic equation $Q=0$,
i.e.
\be
3\lambda^4 -12\lambda^3 +32\lambda^2 +12\lambda -4=0\, .
\mlab{char2}
\ee
Numerical calculations give
$\lambda_c=0.2172$ and
\[
\alpha_1(\lambda_c) = \alpha_2(\lambda_c)
\equiv u_c=0.4435  \, ,\hspace{1cm}
\alpha_3=-\lambda_c-2u_c=-1.1042  \, .
\]

The above value of $\lambda_c$ implies $\alpha_c=0.9098$.

Near the critical value $\lambda \sim \lambda_c$, the imaginary part of
the roots behaves like
\[
v(\lambda )\sim  \sqrt{\lambda -\lambda_c}\, ,
\]
while $\delta (\lambda )\sim v$. Finally we obtain an expression for the
dynamical mass:
\be
\mlab{magg}
m=\Lambda \exp \left[ -\frac{\sigma_1}{\sqrt{\lambda -\lambda_c}}
+\sigma_2 \right]\, ,
\ee
where the constants $\sigma_i$ are given by
\ba
\sigma_1&=&-
\frac{\pi}{2}
\frac{p(\lambda_c)}
{[3Q'(\lambda_c)]^{\frac{1}{2}}}
\\
\sigma_2&=&\half\left[ \frac{1}{u_c}+\frac{1}{1+\lambda_c+u_c}
-2\gamma  \right. \\
&&\left. \phantom{\frac{1}{u_c}}
+\psi (\lambda_c+3u_c)
-\psi (1+\lambda_c+u_c)
-\psi (2+u_c)-\psi (u_c)\right] \, .\nonumber
\ea
Here $\psi (x)=d\log \Gamma (x)/dx$, $\gamma$ is Euler's constant.
Numerically, we find
\be
\mlab{sigg}
\sigma_1=1.525
\hspace{1.5cm}
\sigma_2=1.604  \, .
\ee

\section{Fourth-order Equation}

Our third approximation to the Curtis-Pennington contribution leads to
the following fourth-order differential equation:
\be
\frac{\lambda}{2}\phi (x)f''''+
\left[ 1+\lambda\frac{x}{x+1} \right] xf'''+
\left[ 3-\frac{\lambda}{2}\frac{x^2}{(x+1)^2} \right] f''+
\frac{\lambda}{(x+1)^2}
\left[ f' -f \right] =0 \, ,
\ee
where
\[
\phi (x)=\int_0^xdy\frac{y^2}{y+1}=\half x^2-x+\log (x+1)
\hspace{1mm}\sim \hspace{1mm}\half x^2  \, .
\]
\begin{picture}(0,0)(0.5,-1)
\put(100,3.5){$\scriptscriptstyle x>>1$}
\end{picture}

Substituting $f\sim x^{-\alpha}$, we obtain the characteristic equation
\be
\lambda \alpha^4-2(2-\lambda )\alpha^3-3\lambda\alpha^2
+4(1-\lambda )\alpha -4\lambda =0\, ,
\mlab{char4}
\ee
with the critical value $\lambda_c=0.23320$ and the roots
\[
\alpha_1=\alpha_2=0.4851
\]
\[
\alpha_3=-1.1113
\]
\[
\alpha_4=15.294 \, .
\]
These $\alpha$'s and also $\sigma_1$ can be calculated using the explicit
form for the roots of \mref{char4}.

\chapter{Discussion}

A comparison of the results in the third- and fourth-order cases
with those in the second-order case [\mref{mag2}]
shows that the numerical changes are rather small
(recall that $\alpha_c=\frac{4\pi \lambda_c}{3}$)\footnote{These numbers
for $\alpha_c$ are in agreement with very recent numerical results of
Curtis and Pennington\cite{Mikey}}:

\begin{center}
\begin{tabular}{||c||c|c|c|c||}
\hline
&$\alpha_c$ & $u_c$ & $\sigma_1$& $\sigma_2$\\ \hline
2nd order&1.047&0.5&1.571&1.386\\
3rd order&0.910&0.443&1.525&1.604\\
4th order&0.980&0.485&1.462&\\
\hline
\end{tabular}
\end{center}

We propose to show now that the form \mref{magg} is generic; that is, at
any order of the quenched approximation,
the same asymptotic formula is valid, with
suitable values of the parameters $\lambda_c ,\sigma_1 ,$ and
$\sigma_2$.
Holdom's \cite{Holdom} essentially diagrammatic analysis,
within the quenched approximation,
leads to a bifurcation equation for the mass function of the following
form
\be
f(x)=\lambda \int_0^{\frac{\Lambda^2}{m^2}} dy\frac{yf(y)}{y+1}
K[x,y;0]\, ,     \mlab{Kint}
\ee
where the kernel $K[x,y;0]$ is symmetric under $x \leftrightarrow y$ and
$xK[x,y;0]$ is a function of $x/y$ only. This is known to be in agreement with
a
perturbative expansion to order $O(\alpha^2)$\cite{Appelquist}, and it leads
immediately to a dynamical mass of the form \mref{magg}.

Here we have shown that the same result also holds when we use a
non-perturbative Ansatz for the vertex based on the Ward identity and
multiplicative renormalizability, although this vertex does {\it not}
lead to a bifurcation equation of the form \mref{Kint}. In general this
result will hold for all approximation schemes in which the
integral equation for the mass can be reduced to a differential
equation of the Fuchsian type.

Once we have a differential equation, possibly of very high order,
it is easy to understand that the behaviour of the mass
function at large momenta is $x^{-\alpha}$,
where $\alpha$ is a root of
an (algebraic) characteristic equation of finite order (say $n$),
 with real coefficients.
The onset of oscillations corresponds to equality of two of the roots of
this characteristic equation: for larger values of the coupling, the two
roots are complex conjugate and oscillatory behaviour is manifested and the
ultra-violet conditions can be satisfied.
To calculate the imaginary part $v$ of these roots, let us
consider the characteristic
equation's discriminant:
\[
\Delta = a_0^{2n-2}\prod_{1\le i < j \le n}(\alpha_i-\alpha_j)^2\, ,
\]
($a_0$ is the coefficient of the largest power, $\alpha^n$). The discriminant
is symmetric with respect to the roots $\alpha_i$, and it may be expressed in
terms of the coefficients of the characteristic equation (for the rules to
calculate the discriminant, see for example \cite{discriminant}).
Thus $\Delta$ is an explicit function of $\lambda$.
Note that $\Delta$ changes sign when a pair of complex conjugate roots
appears.

The critical coupling constant $\lambda_c$ is determined by the
condition
\be
\Delta (\lambda_c)=0  \mlab{Delta}
\ee
and the imaginary part $v$ vanishes
like $\Delta^{\frac{1}{2}}$. Normally $\lambda_c$ is a
simple root of \mref{Delta}, so that
\[
v\sim \sqrt{-{\Delta}' (\lambda_c)}\sqrt{\lambda -\lambda_c} , \,
\]
and this leads to formula \mref{magg} for the dynamical mass. In
principle $\lambda_c$ could be a degenerate root, but this would be an
`accident': the generic case is as just given.

Our considerations show that the dependence
of the dynamical
mass on the coupling constant, \mref{magg},  is a universal feature of quenched
theories. The behaviour of the real part of those roots, $u$, is however not
universal. It gives the asymptotic power-envelope of the decreasing and
oscillating mass function:
\[
m(p^2)\sim \left[\frac{1}{p^2}\right]^u=
\left[\frac{1}{p}\right]^{2-\gamma_m}
\, ,
\]
where $\gamma_m$ is the anomalous dimension of the composite operator
$\overline{\psi}\psi$. In the lowest approximation
(the second-order equation), we
have $\gamma_m =1$, but the Curtis-Pennington Ansatz leads to a
modification of this dimension. For example, in the cubic approximation
we find  $\gamma_m =1.114$. Values
$\gamma_m \ge 1$ imply that some four-fermion operators like
$(\overline{\psi}\psi )^2$, $(\overline{\psi}i\gamma_5\psi )^2$, become
relevant in the continuum limit and must be added to the
QED Lagrangian\cite{Bardeen},
which means that pure
QED is not self-consistent.

\section*{Acknowledgements}

We would like to thank V.A. Miransky, M.R. Pennington and F. Wagener
for useful discussions.
Two of us (VPG and PM) wish to thank the Stichting FOM
(Fundamenteel Onderzoek der Materie),
financially supported by the Nederlandse Organisatie voor
Wetenschappelijk Onderzoek, for its support.


\begin{thebibliography}{99}
\bibitem{Maskawa} T. Maskawa and H. Nakajima, Prog. Theor. Phys. {\bf
52} (1974) 1326.
\bibitem{Valery} P.I. Fomin, V.P. Gusynin, V.A. Miransky and Yu.A. Sitenko,
Riv. Nuovo Cim. {\bf 6} (1983) 1.
\bibitem{Valery1} P.I. Fomin, V.P. Gusynin and V.A. Miransky, Phys. Lett.
{\bf B78} (1978) 136;
V.A. Miransky, Nuovo Cim. {\bf 90A} (1985) 149.
\bibitem{Bardeen} W.A. Bardeen, C. Leung and S. Love, Phys. Rev. Lett.
{\bf 56} (1986) 1230; Nucl. Phys. {\bf B273} (1986) 649.
\bibitem{CP} D.C. Curtis and M.R. Pennington,
Phys. Rev. {\bf D42} (1990) 4165; {\bf D46} (1992) 2663.
\bibitem{Kondo} J. Oliensis and P.W. Johnson, Phys. Rev.
{\bf D42} (1990) 656;
V. Gusynin, Mod. Phys. Lett. {\bf A5} (1990)133;
K. Kondo and H. Nakatani, Nucl. Phys. {\bf B351} (1990) 236.
\bibitem{Blatt} D. Atkinson and D.W.E. Blatt, Nucl. Phys. {\bf B151} (1979)
342.
\bibitem{Pimbley} G.H. Pimbley, ``Eigenfunction branches of nonlinear
operators, and their bifurcations'', Lecture Notes in Mathematics, Vol.
104 (Springer, Berlin, 1969);
see also D. Atkinson, J. Math. Phys. {\bf 28} (1987) 2494.
\bibitem{Valery2} V.A. Miransky, Phys. Lett. {\bf B165} (1985) 401;
V.P. Gusynin and V.A. Miransky, Phys. Lett. {\bf B191} (1987) 141.
\bibitem{Fred} D. Atkinson and P.W. Johnson, Phys. Rev. {\bf D35} (1987)
1943.
\bibitem{Kogut} J. Kogut, E. Dagotto and A. Koci\'{c}, Phys. Rev. Lett.
{\bf 59} (1988) 772; Nucl. Phys. {\bf B317} (1989) 253, 271;
Phys. Rev. {\bf D43} (1991) R1763.
\bibitem{Lane} K. Lane, Phys. Rev. {\bf D10} (1974) 2605;
H. Politzer, Nucl. Phys. {\bf B 117}(1976) 397;
V.A. Miransky, Sov. J. Nucl. Phys. {\bf 38} (1984) 280;
K. Higashijima, Phys. Rev. {\bf D29} (1984) 1228.
\bibitem{Yamawaki} T. Akiba and T. Yanagida, Phys. Lett. {\bf B169} 432;
K. Yamawaki, M. Bando and K. Matumoto, Phys. Rev. Lett. {\bf 56} (1986)
1335;
M. Bando, T. Morozumi, H. So and K. Yamawaki,
Phys. Rev. Lett. {\bf 59} (1987) 389.
\bibitem{Holdom} B. Holdom, Phys. Rev. Lett. {\bf 62} (1989) 997.
\bibitem{Mahanta} U. Mahanta, Phys. Lett. {\bf B225} (1989) 187.
\bibitem{Norlund} N.E. N\o rlund, ``Hypergeometric Functions'',
Acta Mathematica {\bf 94} (1955) 289.
\bibitem{Mikey} Private communication from M.R. Pennington.
\bibitem{Appelquist} T. Appelquist, K. Lane and U. Mahanta, Phys. Rev.
Lett. {\bf 61} (1988) 1553.
\bibitem{discriminant} G.A. Korn and T.M. Korn, ``Mathematical Handbook
for Scientists and Engineers'', McGraw-Hill (1968).



\end{thebibliography}
\end{document}